\documentclass[12pt]{article}
\newcommand{\ba}{\begin{array}}
\newcommand{\ea}{\end{array}}
\newcommand{\be}{\begin{equation}}
\newcommand{\ee}{\end{equation}}
\newcommand{\nn}{\nonumber}
\newcommand{\bea}{\begin{eqnarray}}
\newcommand{\ena}{\end{eqnarray}}
\newcommand{\beas}{\begin{eqnarray*}}
\newcommand{\enas}{\end{eqnarray*}}
\newcommand{\mb}{\mbox}

\begin{document}
\pagestyle{empty}

\unitlength=5pt
\newsavebox{\blokht}

\sbox{\blokht}{\begin{picture}(12,12) \put(3,0){\vector(1,0){6}}
\put(3,0){\vector(-1,2){3}} \put(3,12){\vector(-1,-2){3}}
\put(3,12){\vector(1,0){6}} \put(12,6){\vector(-1,-2){3}}
\put(12,6){\vector(-1,2){3}}
\end{picture}}

\newsavebox{\blokhto}

\sbox{\blokhto}{\begin{picture}(12,12) \put(9,0){\vector(-1,0){6}}
\put(0,6){\vector(1,-2){3}} \put(0,6){\vector(1,2){3}}
\put(9,12){\vector(-1,0){6}} \put(9,0){\vector(1,2){3}}
\put(9,12){\vector(1,-2){3}}
\end{picture}}

\newsavebox{\bloktq}

\sbox{\bloktq}{\begin{picture}(6,12) \put(3,0){\vector(-1,2){3}}
\put(3,0){\vector(1,2){3}} \put(3,12){\vector(-1,-2){3}}
\put(3,12){\vector(1,-2){3}}
\end{picture}}

\newsavebox{\bloktqt}

\sbox{\bloktqt}{\begin{picture}(6,12) \put(0,6){\vector(1,-2){3}}
\put(0,6){\vector(1,2){3}} \put(6,6){\vector(-1,-2){3}}
\put(6,6){\vector(-1,2){3}}
\end{picture}}
\begin{center}
\textsf{\Large {An Integrable Model with a non-reducible three
particle R-Matrix }}

\vspace{36pt} {\bf J.~Ambj\o rn\footnote{e-mail:{\sl
ambjorn@alf.nbi.dk}}},\\
\emph{ Niels Bohr Institute, Blegdamsvej 17, Copenhagen, Denmark\\
and \\
Institute for Theoretical Physics, Utrecht University, \\
Leuvenlann 4, 3584 CE Utrecht, The Netherlands }

\vspace{30pt}

 {\bf Sh.~Khachatryan\footnote{e-mail:{\sl
shah@moon.yerphi.am}}}, {\bf A.~Sedrakyan
\footnote{e-mail:{\sl sedrak@jerewan1.yerphi.am}}},\\
\emph{ Yerevan Physics Institute, Alikhanyan Br. str. 2, Yerevan
36, Armenia}
\vfill {\bf Abstract}
\end{center}

We define an integrable lattice model which, in the notation of Yang, in
addition to the conventional 2-particle $R$-matrices also contains
non-reducible 3-particle $R$-matrices. The corresponding modified
Yang-Baxter equations are solved and an expression for the transfer
matrix is found as a normal ordered exponential of a
(non-local) Hamiltonian.

\vfill \rightline{} \rightline{} \rightline{March 2004}

\newpage
\pagestyle{plain} \setcounter{page}{1}

\section{\large Introduction}
\indent

It is well known that integrable models have the following factorization
property: any multi-particle scattering amplitude can be
reduced to the product of 2-particle scattering amplitudes. The
independence of the order in which the successive 2-particle scatterings
take place leads to the Zamolodchikovs so-called triangle equations
for the S-matrix \cite{Z}. In
statistical physics, in the context of integrable lattice systems,
the same property was analyzed earlier by C.N. Yang \cite{Y} and
R.J. Baxter \cite{Bax} using the
so-called $R$-matrices and the equations encounted
were denoted  Yang-Baxter equations (YBE) by
L.Faddeev \cite{F}.

>From the $R$-matrices one can construct the transfer matrices of the
integrable lattice systems and the Yang-Baxter equations impose
sufficient conditions on the $R$-matrices to ensure that transfer
matrices $\tau(u)$ and $\tau(v)$ with different so-called spectral
parameters $u$ and $v$ commute.

In the analysis of the Zamolodchikovs the factorization property
was a consequence of an assumed Lorentz invariance in the model.
Thus it is natural to ask if it is possible to construct an
integrable lattice model with a commuting family of transfer
matrices $\tau(u)$, where the transfer matrices are not products
of only 2-particle $R$-matrices, as is the case for the presently
known integrable models. An obvious, interesting  equation is then
if  such lattice models can be associated with a Lorentz invariant
continuum theory.

In this article we will address this problem and will analyze the
possibility of construction of an integrable model with
3-particle $R$-matrices which can not be reduced to  products of
three 2-particle $R$-matrices. In a previous article \cite{AKhS}
we have considered a model which appeared naturally as a simplification
of the so-called sign-factor representation of the three-dimensional Ising
model (3DIM) on a dual body centered cubic (DBBC) lattice
\footnote{This representation of the Ising model is in many respects
the natural generalization of the representation of the two-dimensional
Ising model as a sum over random walks. Here it becomes a sum over certain
random surfaces, which can be related to fermionic strings.}.
A central point in the construction is the appearance of a two-dimensional
{\it random} Manhattan lattice (ML). Such a Manhattan lattice appears also
in the sign-factor model (SFM) of the 3DIM formulated on a cubic lattice
\cite{AS2},where the hopping of fermions from site to site is allowed only
along the ML arrows.
This directed hopping model can be described as a product of  2-particle
$R$-matrices. However, in the case where the surfaces were
embedded in a DBCC lattice we discovered  the presence of
3-particle $R$-matrices which can not be reduced to the product of
two particle $R$-matrices. Graphically they could be
represented as  honeycombs  \cite{SA},
see Fig 1b.

In this article we go further and construct an integrable model
where the transfer matrix
is a product of 2-particle $R_2$-matrices and (non-reducible) 3-particle
$R_3$-matrices. The model is defined in Section 2.
In Section 3 we show that the corresponding Yang-Baxter equations
ensure that  transfer matrices
with different spectral parameters commute and
we present a  nontrivial solution.
We also obtain an explicit  representation of the transfer matrix
as the normal ordered exponential of a (non-local) Hamiltonian.

\section{ \large Formulation of the model and  definition of the
$R$-matrices}
\indent

 In  \cite{AS2}  the SFM for the 3DIM was defined as a theory of
  fermions interacting with an induced $Z(2)$ gauge field  on
  2d random surfaces embedded in  a three-dimensional regular cubic lattice.
Fermions were hopping  along the directed links of a
certain random ML.
In \cite{AKhS}   we
generalized this construction of a SFM to the case of a 3DIM on a
 DBCC lattice and encountered, as mentioned above, not only
the ordinary 2-particle R-matrices but also 3-particle R-matrices
which could not be factorized into the product of three
2-particle R-matrices.
We initiated the analysis of the SFM by considering the
restriction of the model to a certain regular ML lattice
and investigated it's  integrability.

In this article  we study in the same spirit a model on a more
complicated regular ML. Also this  model has
3-particle $R_3$-
 and ordinary  2-particle $R_2$-matrices,
but due to the different structure of the ML their appearance
in the transfer matrix is different
and we will show that the corresponding  $YBE$'s,
unlike the case considered in \cite{AKhS}, have spectral parameter
dependent solutions.

  The transfer matrix of the model is constructed
  as the product of two monodromy matrices $T_1$ and $T_2$:
  \be \label{Z}
 \tau=tr T_1 T_2,
  \ee
  where  the  matrices  $T_i$  are defined as shown in Fig.2
in terms of the $R$-matrices
  \bea
\label{TT}
 T_1=\prod_n R_3(n) R_2(n) \bar{R}_2(n),\\ \nn
 T_2=\prod_n \bar{R}_2(n)R_2(n)\bar{R_3}(n).
\ena
The $T$-matrices are acting on isomorphic quantum spaces which are
product of two-dimensional spin-spaces
at the horizontal sites 1(n), 2(n), 3(n), 4(n),...
in Fig.2, and they also act on  isomorphic
auxiliary spaces constructed from the left and right boundary
sites in Fig.2. The trace in (\ref{Z}) is taken with respect to the
auxiliary spaces.

Following the technique of  \cite{AKhS} one can express
graphically the $R_2$- and  $R_3$-matrices as shown in Fig.1 and
the transfer matrix $\tau$ as shown in Fig.2. Alternatively
this model can be formulated as an 2d
quantum field model on a ML which can be obtained by repeating
the two rows in Fig.2 in a vertical direction. This lattice is invariant under
translations of the block of two $R_3$ and four $R_2$ matrices in
horizontal direction as well as in  time direction.
Later we will show
that the partition function of the model, $Z=tr (\tau)^{N_0}$,  can be represented
as a functional integral over scalar fermions as
 \be
Z=tr (\tau)^{N_0} = \int D\bar{\psi}D\psi e^{\bar{\psi}
\mathcal{A} \psi + \psi \bar\psi}, \label{ZA}
 \ee
 where the matrix $A_{ij}$   defines the hopping parameters along arrows
on the corresponding ML.
The matrix $A$ inherits the translational invariance of the associated ML.
The   equivalence between the formulation in terms of scalar
fermions as alluded to on the rhs of eq.\ (\ref{ZA}) and the
formulation in terms of the $R$-matrices and $\tau$
is  demonstrated by passing to
a coherent fermionic state basis \cite{coh}
and using the technique developed in \cite{AS4, AK} and will be given below.

 Let us recall the definition of the $R$-matrices.
 The two-particle $(R_2)_{ij}$-matrix is an operator acting on
the direct product of two two-dimensional spaces $V_i$ and $V_j$
with basis elements $|i_1\rangle$ and $|j_1\rangle$, respectively, as
 \be
\label{R2}
 (R_2)_{ij}|i_1\rangle\otimes|j_1\rangle=(-1)^{p(i_2)(p(j_1)+p(j_2))}
 (R_{ij})^{i_2 j_2}_{i_1 j_1}|
 i_2\rangle\otimes|j_2\rangle
 \ee
and can be represented graphically as shown in Fig.1a.

Similarly, the three-particle $(R_3)_{ijk}$-matrix is acting on the direct
product of three two-dimensional spaces $V_i$ , $V_j$ and $V_k$
with basis elements $|i_1\rangle$ , $|j_1\rangle$ and $|k_1\rangle$
 \be
\label{R3}
 (R_3)_{ijk}|i_1\rangle\otimes|j_1\rangle\otimes|k_1\rangle=(-1)^{(p(i_2)+p(j_2))
 (p(k_1)+p(k_2))+p(i_2)p(j_2)}(R_{ijk})^{i_2 j_2 k_2}_{i_1 j_1 k_1}
 |i_2\rangle\otimes|j_2\rangle\otimes|k_2\rangle
 \ee
and can be represented graphically as in Fig.1b.
In the expressions (\ref{R2}) and (\ref{R3})
the parity factor $p(\alpha)$  takes into account
the graded character of the elements
$|\alpha\rangle$ of the Fock space of  scalar fermions
 \be
 p(\alpha)=n,\hspace{2.5cm} |\alpha\rangle=\prod_{i=1}^n c^+_{i}|0\rangle.
 \ee
The basic states of the vector spaces $V_I$ on each
site $i$ are   $|k\rangle$,   $k=0,1$ with
$c|0\rangle=0, |1\rangle=c^+|0\rangle$.

\begin{figure}[t]
\begin{picture}(100,14)
\put(5,8.5){$R_2(p,r)_{12}$}
\multiput(15.5,8.2)(0,0.8){2}{\line(1,0){1}}
\put(22,3){\usebox{\bloktq}}
\multiput(21,9)(0.4,0){20}{\line(1,0){0.1}}\put(17.5,8.5){1,\scriptsize{p}}
\multiput(25,2)(0,0.4){35}{\line(0,1){0.1}}\put(21,1){2,\scriptsize{r}}
\put(37,8.5){$R_3(p,r,s)_{123}$}
\multiput(52,8.2)(0,0.8){2}{\line(1,0){1}}\put(57,3){\usebox{\blokht}}
\multiput(56,9)(0.4,0){37}{\line(1,0){0.1}}\put(54,9){1,\scriptsize{p}}
\multiput(60,2)(0,0.4){35}{\line(0,1){0.1}}\put(57.5,1){2,\scriptsize{r}}
\multiput(66,2)(0,0.4){35}{\line(0,1){0.1}}\put(63,1){3,\scriptsize{s}}

\put(35,-1){Fig.1}
\end{picture}
\vspace{0.5cm}
\end{figure}

In order to have a simple hopping model on the ML
 for both $R_2$ and  $R_3$  we follow the
construction of corresponding matrices
in the SFM of the 3DIM on a DBBC lattice  \cite{AKhS}  and
use as an ansatz  an
 exponential
of a quadratic form (see (\ref{ZA}))
of fermionic creation-annihilation operators:
 \be \label{A}
R_l = :\exp{c^{+}_i A^{(l)}_{ij}c_j}: \qquad l= 2,3, \ee where the
notion $:(\cdot):$ means normal ordering for the odd $i,j$ and
anti-normal(hole) ordering for the even $i, j$. The matrix
elements $(A^{(l)}_{ij}-\delta_{ij}),\; l=2,3$ coincide up
to a sign with the hopping parameters $\mathcal{A}_{ij}$ in
(\ref{ZA})
\be \mathcal{A}_{ij}=\ba{ll}A^{(l)}_{ij}-\delta_{ij},&i -
\mb{odd},\quad j - \mb{even, odd},
\\\delta_{ij}-A^{(l)}_{ij},&i - \mb{even},\quad j - \mb{even, odd}.\ea\ee

 It is straightforward to calculate the matrix elements of
$R_l$ from (\ref{A}) in the Fock space basic. In particular we
note that the fermion number is conserved: \bea \label{pnc}
(R_3)_{ijk}^{\bar{i}\bar{j}\bar{k}}\neq0,\hspace{5mm}
\mbox{if}\hspace{5mm}
i+j+k=\bar{i}+\bar{j}+\bar{k},\hspace{5mm} i,\bar{i},...=0,1, \\
\nn (R_2)_{ij}^{\bar{i}\bar{j}}\neq0,
\hspace{5mm}\mbox{if}\hspace{5mm} i+j=\bar{i}+\bar{j},\hspace{5mm}
i,\bar{i},...=0,1,\
\ena

 \be \label{R33}
 R_3=\left(\ba{cccccccc}
 R_{000}^{000}& & & & & & & \\
  &R_{001}^{001}&R_{010}^{001}& &R_{100}^{001}& & & \\
  &R_{001}^{010}&R_{010}^{010}& &R_{100}^{010}& & & \\
  & & &R_{011}^{011}& &R_{101}^{011}&R_{110}^{011}& \\
  &R_{001}^{100}&R_{010}^{100}& &R_{100}^{100}& & & \\
  & & &R_{011}^{101}& &R_{101}^{101}&R_{110}^{101}& \\
  & & &R_{011}^{110}& &R_{101}^{110}&R_{110}^{110}& \\
  & & & & & & &R_{111}^{111}
 \ea\right)
 \ee
 \be
R_2= \left(\ba{cccc}
R_{00}^{00}& & & \\
 &R_{01}^{01}&R_{01}^{01}& \\
 &R_{01}^{10}&R_{10}^{10}& \\
 & & &R_{11}^{11}
 \ea \right)\ee

The  matrix elements
$(R^{(3)})_{ijk}^{\bar{i}\bar{j}\bar{k}}$ in the expression
(\ref{R33}) are connected with the $A_{ij}$'s in (\ref{A}) by the
following equations:
\bea \nn \label{1RA} \hspace{-25mm}
\begin{array}{ll}
R_{000}^{000}=(1+A^{(3)}_{11})(1+A^{(3)}_{33})-A^{(3)}_{13}A^{(3)}_{31},&\hspace{11mm}R_{101}^{101}=1,\\
R_{011}^{011}=(1+A^{(3)}_{11})(1-A^{(3)}_{22})+A^{(3)}_{12}A^{(3)}_{21},&\hspace{11mm}R_{001}^{001}=1+A^{(3)}_{11},\\
R_{110}^{110}=(1+A^{(3)}_{33})(1-A^{(3)}_{22})+A^{(3)}_{23}A^{(3)}_{32},&\hspace{11mm}R_{111}^{111}=1-A^{(3)}_{22},\\
R_{010}^{010}=-\det
(A^{(3)}-1),&\hspace{11mm}R_{100}^{100}=1+A^{(3)}_{33},\\
\end{array}
\ena

\be\nn \label{2RA} \hspace{-20mm}
\begin{array}{ll}
R_{001}^{010}=A^{(3)}_{23}(1+A^{(3)}_{11})-A^{(3)}_{13}A^{(3)}_{21},&\hspace{11mm}R_{010}^{001}=
A^{(3)}_{32}(1+A^{(3)}_{11})-A^{(3)}_{31}A^{(3)}_{12},\\
R_{010}^{100}=A^{(3)}_{12}(1+A^{(3)}_{33})-A^{(3)}_{32}A^{(3)}_{13},&\hspace{11mm}R_{100}^{010}=
A^{(3)}_{21} (1+A^{(3)}_{33})-A^{(3)}_{23}A^{(3)}_{31},\\R_{011}^{110}=A^{(3)}_{13}(1-
A^{(3)}_{22})+A^{(3)}_{12}A^{(3)}_{23},&\hspace{11mm}R_{110}^{011}=A^{(3)}_{31}(1-A^{(3)}_{22})+A^{(3)}_{21}A^{(3)}_{32},\\
R_{001}^{100}=-A^{(3)}_{13}&\hspace{11mm}R_{100}^{001}=-A^{(3)}_{31},\\
R_{101}^{110}=A^{(3)}_{23},&\hspace{11mm} R_{110}^{101}=A^{(3)}_{32},\\
R_{011}^{101}=A^{(3)}_{12},&\hspace{11mm}R_{101}^{011}=A^{(3)}_{21}.\\
\end{array}
\label{RA} \ee
 The two-particle $R_2$-matrix elements
  can be obtained from these expressions by
taking  $A^{(3)}_{i3}=A^{(3)}_{3j}=0$
everywhere.

There are also some additional model
dependent constraints for the $R_3$ matrix para\-meters
coming from the specific  lattice used:
 \be
 A^{(3)}_{22}-1=A^{(3)}_{13}=A^{(3)}_{31}=0,
 \ee
These conditions express  algebraically that we
(by the lattice contruction) have no hoppings across
the hexagon (see Fig.2), and this is the reason the
$R_3$-matrix can not be reduced to the product of three $R_2$'s,
as one might have expected from general  factorization properties.
Going furter back to the original full SFM on a random ML coming
from the 3DIM, the hexagon $R_3$-matrices are precisely
associated with part of the embedded surfaces
which carry a curvature, unlike the square
$R_2$ matrices which are associated with flat parts of
the embedded surfaces. In the piecewise linear geometry it is well known
that the curvature
\begin{equation}
\label{RR}
K_n = \frac{\pi}{4}(4 - n )
\end{equation}
is associated with the n-faces of two dimensional complexes.
Therefore $K_6= -\pi/2$, while $K_4= 0$.

Viewing the R-matrices as
associated with particle scattering one
attaches, using the language of integrable systems, a
spectral parameter to each of the particles, the spectral
parameters being connected to the rapidities of the particles..
Therefore one expects that the $R_3$-matrix in general
depends on three spectral
parameters $(p,r,s)$,  while $R_2$  depends on  two spectral
parameters $(p,r)$.

\section{ \large Yang-Baxter equations and their solutions }
\indent

The integrability conditions for the
model can be found in the standard way
by defining and solving the associated YBEs
equations.  One can obtain the local YBEs using the graphical
representation  shown in  Fig.\ 3a and Fig.\ 3b as described
for instance in \cite{F}:
 \be
 \overrightarrow{R}_{12}(p,q) R_{234}(p,r,s)R_{12}(q,r)
\bar{R}_{34}(q,s)
 =\bar{R}_{234}(q,r,s)\bar{R}_{23}(p,r)R_{34}(p,s)
\overrightarrow{R}_{34}(p,q),
 \label{eq3}
\ee
\be \label{eq3*}
\overrightarrow{R}_{12}(p,q)\bar{R}_{23}(p,r)R_{34}(p,s)
\bar{R}_{123}(q,r,s)=
R_{12}(q,r)\bar{R}_{34}(q,s)R_{234}(p,r,s)\overleftarrow{R}_{34}(p,q),
\ee
These equations
 ensure that the
transfer matrices $\tau(p,r,s)=Tr (T_1 T_2)$
($T_1$ and $T_2$ are defined by (\ref{TT})) with different
spectral parameters commute:
\be \label{CM}
[\tau(p,r,s),\tau(q,r,s)]=0.
\ee
(we have just written the commutativity condition for one of spectral
paramenters).
This is
what we mean by the system being integrable since
we can use (\ref{CM}) to define an infinite
set of mutually commuting, conserved charges
$H_n(r,s)$ by expanding the transfer
matrix  $\tau(p,r,s)$ in powers of p:
 \be \label{Hn}
 H_n(r,s)=\frac{d^n T(p,r,s)}{dp^n}|_{p=p_0}.
 \ee

 We now look for non-trivial solutions
to (\ref{eq3}) and (\ref{eq3*}) where the intertwiner matrices
 $\overleftarrow{R}_{12}(p,q),
 \overrightarrow{R}_{34}(p,q)$ have the  structure (\ref{A}) and (\ref{2RA}).
For convenience let us  use a specific notation for the matrix
elements of the $R_2$- and $\bar{R}_2$-matrices, while keeping
$A^{(3)}_{ij}$'s  for the parameterization of
 $R_3$, $\bar{R}_3$-matrices
\be
\label{S1}
 \ba{ll}
 R_{00}^{00}(p,r)=a_1(p,r),&R_{11}^{11}(p,r)=a_2(p,r),\\
 R_{01}^{10}(p,r)=b_1(p,r),&R_{10}^{01}(p,r)=-b_2(p,r),\\
\bar{R}_{00}^{00}(p,s)=\bar{a}_1(p,s),&\bar{R}_{11}^{11}(p,s)=\bar{a}_2(p,s),\\
\bar{R}_{01}^{10}(p,s)=\bar{b}_1(p,s),&\bar{R}_{10}^{01}(p,s)=-\bar{b}_2(p,s).\\
\ea \ee
With this notation the $YBE$s
(\ref{eq3}) and (\ref{eq3*}) reduce to the following constraints
 on the parameters $a_k, \bar{a}_k, b_k, \bar{b}_k, \qquad k=1,2$.
\bea
\label{S2}
\frac{a_1(p,r)a_2(p,r)}{b_1(p,r)b_2(p,r)} &=& f(r),\hspace{2.5cm}
\frac{\bar{a}_1(p,r)\bar{a}_2(p,r)}{\bar{b}_1(p,r)\bar{b}_2(p,r)} = \bar{f}(s)\\
\nn a_1(p,r)\bar{b}_2(p,s)&=&\alpha_{12}(r,s),
\hspace{2cm}a_2(p,r)\bar{b}_2(p,s)=\alpha_{21}(r,s),\\ \nn
\bar{a}_1(p,s)b_2(p,r)&=&\bar{\alpha}_{12}(r,s),
\hspace{2cm}\bar{a}_2(p,s)b_1(p,r)=\bar{\alpha}_{21}(r,s)
 \ena
and the $R_3$ and $\bar{R}_3$ elements are connected with
$a_i,b_i$  by the relations
\be
\label{S3}
 \ba{ccc}
A^{(3)}_{23}(q,r,s)=y(r,s),&A^{(3)}_{21}(q,r,s)=x(r,s)\bar{a}_1(p,s),&A^{(3)}_{11}(q,r,s)=v(r,s)a_2(p,r)-1,\\
\bar{A}^{(3)}_{23}(p,r,s)=\bar{y}(r,s),&\bar{A}^{(3)}_{21}(p,r,s)=\bar{x}(r,s)a_1(p,r),
&\bar{A}^{(3)}_{11}(p,r,s)=\bar{v}(r,s)\bar{a}_2(p,s)-1,
\\
A^{(3)}_{12}(q,r,s)=w(r,s),&A^{(3)}_{32}(q,r,s)=u(r,s)\bar{b}_1(p,s),&A^{(3)}_{33}(q,r,s)=z(r,s)b_2(p,r)-1,
\\
\bar{A}^{(3)}_{12}(p,r,s)=\bar{w}(r,s),
&\bar{A}^{(3)}_{32}(p,r,s)=\bar{u}(r,s)b_1(p,r),&\bar{A}^{(3)}_{33}(p,r,s)=\bar{z}(r,s)\bar{b}_2(p,s)-1.
\ea \ee

Here  $ f(r), \bar{r}(r), \alpha_{ij}(r,s), \bar{\alpha}_{ij}(r,s),
y(r,s), \bar{y}(r,s),  w(r,s), \bar{w}(r,s), x(r,s), \bar{x}(r,s), \\
 u(r,s), \bar{u}(r,s), v(r,s), \bar{v}(r,s),
z(r,s), \bar{z}(r,s),$  are arbitrary functions of
variables $r,s$.

\vspace{1cm}
\section{\large The Transfer Matrix}
\indent

In order to calculate the transfer matrix of our model explicitly,
expressed  as a normal ordered form of an exponential
operator, one inserts (\ref{A}), given by
(\ref{S1})--(\ref{S3}), into (\ref{TT}) (see Fig.2). From
Fig.\ 2 it follows that the double-row transfer matrix (\ref{TT})
is invariant under
translations by four lattice spacings:  we can
restore the whole lattice
structure by translation of the block of  two $R_3$ - and
four $R_2$-matrices, either horizontally or vertically.
After some algebra, using Wick`s contraction theorem, we obtain:
\be \label{Hprs}
 \tau(p,r,s)=tr(T_1T_2)=F(r,s):\exp{H(p,r,s)}:.
 \ee
The number-valued prefactor $F(r,s)$ of the normal ordered
exponential is
\bea\label{f}
F(r,s)=\lambda_1^N+\lambda_2^N-\varepsilon_1^N-\varepsilon_2^N,
\ena
where $N$ is the number of constituent horizontal
blocks in the chain and
\bea\label{f1} \lambda_{1,2} &=&\frac{\vartheta\pm
\sqrt{\vartheta^2-4\varepsilon_1\varepsilon_2}}{2},\nn\\
\vartheta &=&(1-\bar{\alpha}_{12}\bar{\omega})
(1-\bar{\alpha}_{21}\bar{y})+(1-u\bar{z}\alpha_{12}\alpha_{21})
(1-v\bar{x}\alpha_{12}\alpha_{21})-1,\nn\\
\varepsilon_1&=&\bar{\alpha}_{21}\alpha_{12}^2\bar{x}\bar{z}\omega,\qquad
\varepsilon_2=\bar{\alpha}_{12}\alpha_{21}^2u v\bar{y},\nn \ena

The operator $H(p,r,s)$ in (\ref{f}) is a quadratic form of fermionic
creation and annihilation  operators:
\be
H(p,r,s)=\sum_{i,j}H_{ij}(n-m)c^+_i(n)c_j(m).
\ee
This will be derived below.

In general $H$  is a nonlocal operator,
$H_{ij}(n-m)$ being a polynomial function of  $\lambda_{1,2}$
and $\varepsilon_{1,2}$   of degrees
$i-j$, or $N-(i-j)$, respectively.
But by the translational invariance of the
monodromy matrices we can Fourier transform the
$H(p,r,s)$ operators and present them
in a compact form as
\bea \label{F}
H(p,r,s)&=&\sum_{n,m;i,j}^{N;4}H_{ij}(n-m)c^+_i(n)c_j(m)=
\sum_{i,j;k}^{4;N}\bar{H}_{i,j}(k)c^+_i(k)c_j(-k),\nn\\
c^+_i(k)&=&\frac{1}{\sqrt{N}}\sum_{n=1}^{N}e^{i2\pi\frac{k
n}{N}}c^+_i(n),\qquad \qquad
c_j(k)=\frac{1}{\sqrt{N}}\sum_{n=1}^{N}e^{-i2\pi\frac{k
n}{N}}c_j(n),\nn\\
\bar{H}_{i,j}(k)&=&\frac{1}{N}\sum_{n=1}^{N}e^{i2\pi\frac{k
n}{N}}H_{i,j}(n).\ena

Let us now finally discuss how the
calculation of the matrix elements of $\bar{H}_{i,j}(k)$
can  be done  by using the
technique of fermionic coherent states with Grassmannian
variables, as formulated in \cite{coh, AS4, AS5}.
In this basis the product and the trace of the operators are obtained
by integrating over the Grassmann variables.
 Choosing a anti-coherent and coherent basis
for particles on the odd $(2k-1;n)$ and even $(2k;n)$
boundary sites of the monodromy matrices (\ref{TT}),
respectively, (see Fig.2),  we define
 \be \ba{ll} |\psi
_{2k-1}(n)\rangle=e^{\psi_{2k-1}(n)c^+_{2k-1}(n)}|0\rangle,&
\langle\bar \psi_{2k-1}(n)|=\langle0|e^{c_{2k-1}(n)\bar \psi_{2k-1}(n)} ,\\
|\bar \psi_{2k}(n)\rangle=(c^+_{2k}(n)-\bar
\psi_{2k}(n))|0\rangle,&
\langle\psi _{2k}(n)|=\langle0|(c_{2k}(n)+\psi _{2k}(n)),\\
\langle\bar \psi _{2k-1}(n)|\psi _{2k-1}(n)\rangle=e^{\bar \psi
_{2k-1}(n)\psi _{2k-1}(n)},& \langle\psi _{2k}(n)|\bar\psi
_{2k}(n)\rangle=e^{\psi _{2k}(n)\bar\psi_{2k}(n)}.\\
 \ea
\label{coherent} \ee
These states are by construction eigenstates of
the fermionic creation and annihilation operators  $c_k^+$ and $c_k$
with eigenvalues $\psi_k(n)$ and $\bar\psi_k(n)$
\begin{eqnarray}
\label{CPR}
c_{2k}\mid \psi_{2k}(n)\rangle =- \psi_{2k}(n)\mid \psi_{2k}(n)\rangle &,&
\langle\bar{\psi}_{2k}(n)\mid c^+_{2k} =
-\langle\bar{\psi}_{2k}(n) \mid \bar{\psi}_{2k}(n),\\
c^+_{2k+1}\mid \bar{\psi}_{2k+1}(n)\rangle =
\bar{\psi}_{2k+1}(n)\mid \bar{\psi}_{2k+1}(n)\rangle &,&
\langle\psi_{2k+1}(n)\mid c_{2k+1} =- \langle\psi_{2k+1}(n)\mid \psi_{2k+1}(n).\nn
\end{eqnarray}

We attach also coherent
states $\chi_{2k}(n), \; \bar\chi_{2k}(n),\; \chi_{2k+1}(n), $
and $ \bar\chi_{2k+1}(n)$
to the intermediate sites between the
two transfer matrices $\tau_1$ and $\tau_2$.
Then the full two row-transfer matrix in the coherent states basis, expressed
via the one-row transfer matrices $\tau_1$ and $\tau_2$, can be written as
\bea
\label{t}
 \tau(\bar\psi,\psi)&=&\int D\bar{\chi} D\chi e^{-\sum \bar{\chi}\chi}
 \tau_1(\bar{\psi}_{2k-1}(n),\psi_{2k}(n),\bar{\chi}_{2k}(n),
\chi_{2k-1}(n))\\
 &&\cdot \tau_2(\bar{\chi}_{2k-1}(n),\chi_{2k}(n),
\bar{\psi}_{2k}(n),\psi_{2k-1}(n)),\nn
\ena
where
\bea
\label{t1}
\tau_i&=&tr T_i,\qquad \qquad i = 1,2 \nn \\
\tau_1(\bar\psi,\psi,\bar\chi,\chi)&=&\prod_{n=1}^{N} \prod_{k=1}^{2}\langle
\bar{\psi}_{2k-1}(n)| \langle \psi_{2k}(n)|
 \tau_1| \bar{\chi}_{2k}(n)\rangle | \chi_{2k-1}(n)\rangle,\nn \\
\tau_2(\bar\chi,\chi,\bar\psi,\psi)&=&\prod_{n=1}^{N} \prod_{k=1}^{2}\langle
\bar{\chi}_{2k-1}(n)| \langle \chi_{2k}(n)|
 \tau_2 | \bar{\psi}_{2k}(n)\rangle | \psi_{2k-1}(n)\rangle.
 \ena

In order to define matrix
multiplication in the  coherent space basis we simply insert
between operators the  identity operators
\bea \label{cohb}
 \int d\bar \chi_{i}(n) d\chi_{i}(n) |\chi _i(n)\rangle\langle\bar
 \chi _i(n)|e^{-\bar \chi_{i}(n)
\chi_{i}(n)}=1, \nn \\
 \int d\bar \chi_{i}(n) d\chi_{i}(n) |\bar{\chi} _i(n)\rangle\langle
 \chi _i(n)|e^{-\bar \chi_{i}(n)
\chi_{i}(n)}=1 \ena
for coherent and anticoherent states, respectively.
These relations simply express the completness of the coherent state basis.

Now it is straightforward to express the
transfer matrix in the coherent states basis. By inserting
expressions (\ref{cohb}) for the intermediate coherent states
$\chi$ between $R$-operators in \ref{TT}
and by considering matrix elements
between external external quantum states $\psi$ we  obtain
\bea
\nn \tau(\bar\psi,\psi)=\int D\bar{\chi} D\chi e^{-\sum
\bar{\chi}\chi}\prod
\textbf{R}_i(n)(\bar{\psi},\psi,\bar{\chi},\chi)=\\\nn \int
D\bar{\chi} D\chi D\bar{\chi} D\chi \exp\{
\sum_{n-m=0,1}(\bar{\chi}_i(n)\Delta_{ij}(n-m)\chi_j(m)+\bar{\psi}_i(n)\bar{\Delta}_{ij}(n-m)
\psi_j(m)\\+ \bar{\psi}_i(n)\bar{\bar{
\Delta}}_{ij}(n-m)\chi_j(m)+\bar{\chi}_i(n)\bar{\bar{\bar{\Delta}}}_{ij}
(n-m)\psi_j(m))\}.
 \ena
In this equation the matrix elements of the operator
$R_l(\bar c,c)$ from  (\ref{A}) are represented as
 $\textbf{R}(\bar{\psi},\psi)=e^{\sum\bar{\psi}\psi}R_l(\bar{\psi},\psi)$
by use of the coherent states.

The matrix $\Delta$ represents vacuum fluctuations
and its determinant will appear in the final expression
after integration over internal states $\chi$.
 After Fourier
 transformation of the Grassmanian variables $\psi,\chi$,
the  Fourier transformed $\Delta(k)$
 of the matrix $\Delta(n)$
 becomes a $10\times 10$ matrix
(10 is the number of the intermediate states in each repeating block,
 as one can see in Fig.2):
 \be
 \Delta(k)=\frac{1}{N}\sum_{n=1}^{N}\Delta(n)=
 \left(
 \ba{llllllllll}
 -1&b_2&0&0&0&0&0&0&a_1e^{2i\pi\frac{k}{N}}&0\\
0&-1&\bar{a}_1&0&0&0&0&0&0&0 \\
w&0&-1&v&0&0&0&0&0&0\\
0&0&0&-1&0&\bar{b}_1&0&0&0&0
\\ 0&0&\bar{b}_2&0&-1&0&0&0&0&0
\\ 0&0&0&0&\bar{x}&-1&0&\bar{y}&0&0
\\ 0&0&0&0&0&\bar{a}_2&-1&0&0&0
\\ 0&0&0&0&0&0&b_1&-1&0&a_2
\\ 0&0&0&0&0&0&0&\bar{z}&-1&0
\\ ue^{-2i\pi\frac{k}{N}}&0&0&0&0&0&0&0&0&-1
 \ea
 \right).
 \ee

The function $F(p,r,s)$ in eq.\ (\ref{Hprs}) can be expressed
via the determinant of $\Delta$ as
  \be
F(p,r,s)= \det{\Delta_{i,j}(n,m)}=\prod_k \det{\Delta_{ij}(k)},
  \ee
while   $H(p,r,s)$ is defined by the Fourier transform of
the matrix
\be
\bar{H}_{i,j}(k) =\bar{\Delta}_{ij}(k)+\bar{\bar{\Delta}}_{im}(k)(\Delta)^{-1}_{ml}(k)\bar{\bar{\bar{\Delta}}}_{lj}(k)
\ee

We will write down here $\bar{H}_{i,j}(k)$ only for a
simple case:
\bea
\nn
 \alpha_{12}=\alpha_{21}=\bar{\alpha}_{12}=\bar{\alpha}_{21}=\alpha,\\\nn
 f=\bar{f}=1,\\
 w=\bar{y}=\bar{w}=y,\\\nn
 u=v=x=z=\bar{u}=\bar{v}=\bar{x}=\bar{z}
 \ena
and have found following  matrix elements of the Hamiltonian
\bea
\bar{H}_{11}(k,\frac{b_1}{a_1})=-A(\frac{b_1}{a_1})^2\alpha^2u^2[\alpha
w e^{i\frac{2\pi k}{N}}-\alpha^2u^2+1]-1
\nn\\
\bar{H}_{11}(k,\frac{b_1}{a_1})=
\bar{H}_{22}(k,\frac{a_1}{b_1})=\bar{H}_{33}(-k,
\frac{b_1}{a_1})
=\bar{H}_{44}(-k,\frac{a_1}{b_1}),\nn \\
\bar{H}_{14}=\bar{H}_{41}=-A u \alpha(1-\alpha^2 u^2-w\alpha)\nn\\
 \bar{H}_{12}(k)=\bar{H}_{43}(-k)=-A\alpha(1+e^{-i\frac{2\pi k}{N}}\alpha^2
u^2-w\alpha)\nn\\
\bar{H}_{23}(k)=\bar{H}_{32}(-k)=-Au \alpha
e^{i\frac{2\pi k}{N}}(1-\alpha^2 u^2-w\alpha)\nn\\
\bar{H}_{21}(k)=\bar{H}_{34}(-k)=-Au^2\alpha^2 w e^{i\frac{2\pi
k}{N}}(1+\alpha^2 u^2e^{-i\frac{2\pi k}{N}}-w\alpha)+w\nn\\
 \bar{H}_{24}(k,\frac{b_1}{a_1})=u\alpha^2
\bar{H}_{42}(-k,\frac{b_1}{a_1})=-A(\frac{a_1}{b_1})^2 u
\alpha^2(e^{i\frac{2\pi
k}{N}}+1)\nn \\
\bar{H}_{31}(k,\frac{b_1}{a_1})=u^2\alpha^2 w
\bar{H}_{13}(-k,\frac{b_1}{a_1})=-A(\frac{b_1}{a_1})^2 u^2\alpha(
e^{i\frac{2\pi k}{N}}+1)\ena
where
\be
\nn A^{-1}=(1-w\alpha)^2+(1-u^2\alpha^2)-1-2\cos{(2\pi
k/N)}u^2w \alpha^3,\nn
\ee

It should be mentioned that the dependence on
 the spectral parameter $p$ in the expressions above
comes from the fraction $a_1/b_1$.
As  seen we have a nonlocal model of hopping fermions.

Finally we would like to make the following remark. In  the
article \cite{AZ} A. Zamolodchikov has defined some $S$-matrix for
scattering of  straight strings and formulated the analog of the
YBEs for them, called the Tetrahedron Equations, in order to have
an integrable model.  In the vertex formulation this $S$-matrix
\cite{BSKor} has three initial and three final indices precisely
as the $R_3$-matrix in our construction. Contrary to our
situation, where particle  number conservation is ensured by eq.\
(\ref{pnc}), the non-trivial (to us known) solutions of the
Tetrahedron Equations do not have particle number conservation,
except of the one case represented in \cite{BS1}.

\section*{\large Acknowledgment}
J.A.\, Sh.K. and A.S.\ were supported by an INTAS grant.
J.A.\   also acknowledges support by the
EU network on ``Discrete Random Geometry'', grant HPRN-CT-1999-00161
as well as  by {\it MaPhySto}, Network of Mathematical Physics
and Stochastics, funded by a grant from Danish National Research Foundation.

\newpage

\unitlength=5pt

\begin{picture}(100,24)
\multiput(0,0)(24,0){3}{\usebox{\blokht}}
\multiput(12,12)(24,0){3}{\usebox{\blokhto}}
\multiput(6,12)(24,0){3}{\usebox{\bloktq}}
\multiput(18,0)(24,0){3}{\usebox{\bloktqt}}
\multiput(0,12)(24,0){3}{\usebox{\bloktqt}}
\multiput(12,0)(24,0){3}{\usebox{\bloktq}} \put(-4,17){$T_1$}
\put(-4,5){$T_2$}
\end{picture}

\begin{picture}(100,2)(0,-2)
\put(2.4,18){$\bar{R}_2$} \put(8.4,18){$R_2$}\put(17.4,18){$R_3$}
\put(5.2,6){$\bar{R}_3$}\put(14.4,6){$R_2$}\put(20.4,6){$\bar{R}_2$}
\put(3,-1.5){$1(n)$} \put(9,-1.5){$2(n)$} \put(15,-1.5){$3(n)$}
\put(21,-1.5){$4(n)$} \put(27,-1.5){$1(n+1)$}
\end{picture}

\begin{picture}(50,5)
\put(34,2){Fig.2}
\end{picture}

\vspace{2cm}
\newsavebox{\blokhi}

\sbox{\blokhi}{\begin{picture}(12,12) \put(3,0){\vector(1,0){6}}
\put(3,0){\vector(-1,2){3}} \put(3,12){\vector(-1,-2){3}}
\put(3,12){\vector(1,0){6}} \put(12,6){\vector(-1,-2){3}}
\put(12,6){\vector(-1,2){3}}
\end{picture}}

\newsavebox{\blokhoi}

\sbox{\blokhoi}{\begin{picture}(12,12) \put(9,0){\vector(-1,0){6}}
\put(0,6){\vector(1,-2){3}} \put(0,6){\vector(1,2){3}}
\put(9,12){\vector(-1,0){6}} \put(9,0){\vector(1,2){3}}
\put(9,12){\vector(1,-2){3}}
\end{picture}}

\newsavebox{\blokqi}

\sbox{\blokqi}{\begin{picture}(6,12) \put(3,0){\vector(-1,2){3}}
\put(3,0){\vector(1,2){3}} \put(3,12){\vector(-1,-2){3}}
\put(3,12){\vector(1,-2){3}}
\end{picture}}

\newsavebox{\blokqti}

\sbox{\blokqti}{\begin{picture}(6,12) \put(0,6){\vector(1,-2){3}}
\put(0,6){\vector(1,2){3}} \put(6,6){\vector(-1,-2){3}}
\put(6,6){\vector(-1,2){3}}
\end{picture}}

\newsavebox{\blokri}

\sbox{\blokri}{\begin{picture}(12,12) \put(0,12){\vector(1,0){12}}
\put(0,12){\vector(0,-1){12}} \put(12,0){\vector(-1,0){12}}
\put(12,0){\vector(0,1){12}}
\end{picture}}

\newsavebox{\blokrti}

\sbox{\blokrti}{\begin{picture}(12,12)
\put(12,12){\vector(-1,0){12}} \put(0,0){\vector(0,1){12}}
\put(0,0){\vector(1,0){12}} \put(12,12){\vector(0,-1){12}}
\end{picture}}

\begin{picture}(40,24)
\put(18,12){\usebox{\blokhoi}} \put(24,0){\usebox{\blokqti}}
\put(18,0){\usebox{\blokqi}}
 \put(6,6){\usebox{\blokri}}
\multiput(34,11.5)(0,1){2}{\line(1,0){1.5}}
\put(9,12){$\overleftarrow{R_2}(p,q)$}
\put(22,18){\scriptsize{$R_3(p)$}}
\put(19.1,6){\scriptsize{$R_2(q)$}}
\put(25.1,6){\scriptsize{$\bar{R}_2(q)$}} \put(5,18.1){1}
\put(5,5){2} \put(22,-1){3} \put(28,-1){4}
\put(21.8,24.2){$\acute{1}$} \put(27.8,24.2){$\acute{2}$}
\put(30.5,18.1){$\acute{3}$} \put(30.5,5){$\acute{4}$}
\end{picture}
\hspace{-1.5cm}
\begin{picture}(50,24)(-1,1)
\put(8,13){\usebox{\blokhoi}} \put(14,1){\usebox{\blokqti}}
\put(8,1){\usebox{\blokqi}} \put(20,7){\usebox{\blokrti}}
\put(24,12){$\overrightarrow{R_2}(p,q)$}
\put(15.5,6.5){\scriptsize{$R_2(p)$}}
\put(9.5,6.5){\scriptsize{$\bar{R}_2(p)$}}
\put(12.5,18.5){\scriptsize{$\bar{R}_3(q)$}} \put(7,18.1){1}
\put(7,5){2} \put(11.5,-1){3} \put(17.5,-1){4}
\put(11.5,25.1){$\acute{1}$} \put(17.5,25){$\acute{2}$}
\put(32,19.1){$\acute{3}$} \put(32,5){$\acute{4}$}
 \put(1.5,-3){Fig.3a}
\end{picture}

\vspace{3cm}

\begin{picture}(40,20)
\put(20,1){\usebox{\blokhoi}} \put(26,13){\usebox{\blokqti}}
\put(20,13){\usebox{\blokqi}} \put(8,7){\usebox{\blokrti}}
\put(11,14){$\overrightarrow{R}_2(p,q)$}
\multiput(35,11.5)(0,1){2}{\line(1,0){1.5}}
\put(8,19.4){1} \put(7,5){2} \put(24,-1){3} \put(30,-1){4}
\put(23.8,24.2){$\acute{1}$} \put(29.8,24.2){$\acute{2}$}
\put(32.5,18.1){$\acute{3}$} \put(32.5,6){$\acute{4}$}
\put(21,18.5){\scriptsize{$\bar{R}_2(p)$}}
\put(27.2,18.5){\scriptsize{$R_2(p)$}}
\put(24.5,6.5){\scriptsize{$\bar{R}_3(q)$}}
 \put(35,-2){Fig.3b}
\end{picture}
\hspace{3cm}
\begin{picture}(50,26)(20,-1)
\put(6,0){\usebox{\blokhoi}} \put(12,12){\usebox{\blokqti}}
\put(6,12){\usebox{\blokqi}}
 \put(18,6){\usebox{\blokri}}
 \put(20,12){$\overleftarrow{R}_2(p,q)$}
\put(7.2,18.5){\scriptsize{$\bar{R}_2(q)$}}
\put(13.5,18.5){\scriptsize{$R_2(q)$}}
\put(10.5,6.5){\scriptsize{$\bar{R}_3(p)$}} \put(5,18.1){1}
\put(5,5){2} \put(10,-1.5){3} \put(16,-1.5){4}
\put(9.8,24.2){$\acute{1}$} \put(15.8,24.2){$\acute{2}$}
\put(30,18.1){$\acute{3}$} \put(30,5){$\acute{4}$}
\end{picture}

\vspace{2cm}

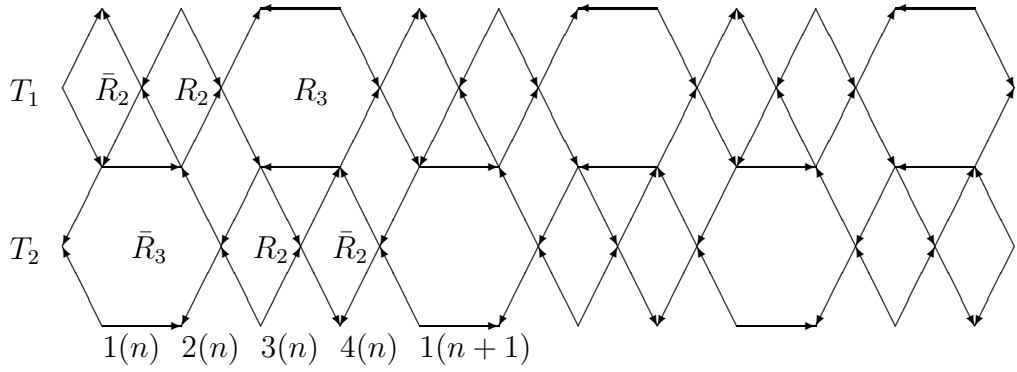
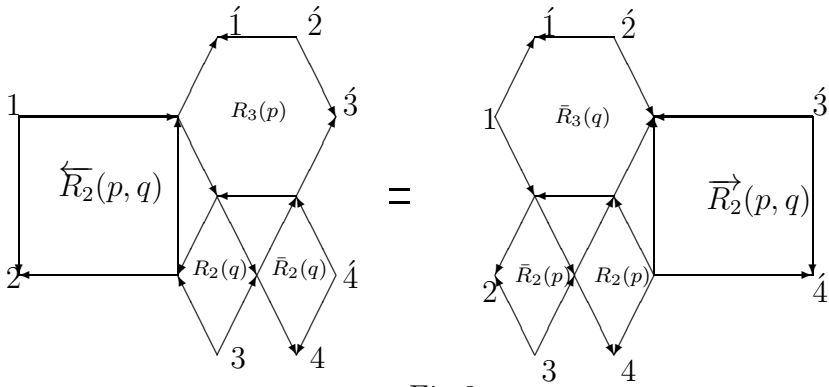
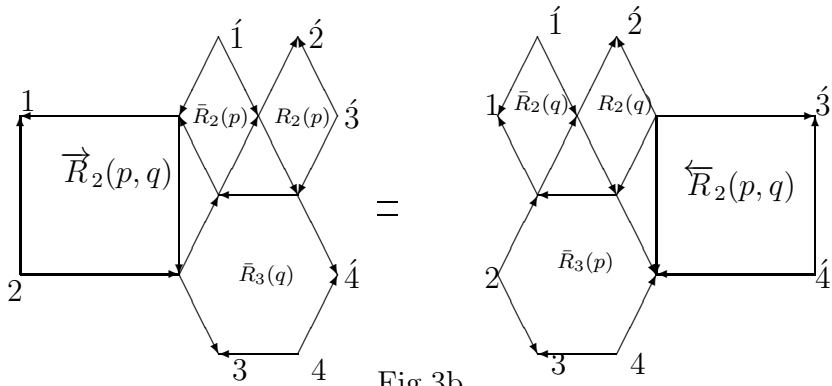
\end{document}